Dominant ferromagnetism in the spin-½ half-twist ladder 334 compounds, $Ba_3Cu_3In_4O_{12}$ and $Ba_3Cu_3Sc_4O_{12}$

S. E. Dutton[1,*], M. Kumar[1], Z. G. Soos[1], C. L. Broholm[2] and R. J. Cava[1]

[1]Department of Chemistry, Princeton University, Princeton, NJ 08544, USA

[2]Institute for Quantum Matter and Department of Physics and Astronomy, The Johns Hopkins University, Baltimore, MD 21218, USA

[*]corresponding author: sdutton@princeton.edu

Abstract

The magnetic properties of polycrystalline samples of $Ba_3Cu_3In_4O_{12}$ (In-334) and $Ba_3Cu_3Sc_4O_{12}$ (Sc-334) are reported. Both 334 phases have a structure derived from perovskite, with $CuO_4$ squares interconnected to form half-twist ladders along the *c*-axis. The Cu-O-Cu angles, ~ 90°, and the positive Weiss temperatures indicate the presence of significant ferromagnetic (FM) interactions along the Cu ladders. At low temperatures, $T < 20$ K, sharp transitions in the magnetic susceptibility and heat capacity measurements indicate three-dimensional (3D) antiferromagnetic (AFM) ordering at $T_N$. $T_N$ is suppressed on application of a field and a complex magnetic phase diagram with three distinct magnetic regimes below the upper critical field can be inferred from our measurements. The magnetic interactions are discussed in relation to a modified spin-½ FM-AFM model and the 334 half-twist ladder is compared to other 2-rung ladder spin-½ systems.

## 1. Introduction

Despite their apparent simplicity,[1] materials containing one-dimensional (1D) spin-chains are frequently complicated by significant nearest neighbor (NN), $J_1$, and next nearest neighbor (NNN), $J_2$, interactions.[2-4] For both ferromagnetic (FM) and antiferromagnetic (AFM) $J_1$ interactions, competition with an AFM $J_2$ can result in multiple low energy magnetic regimes and a complex phase diagram.[5] These models have been realised in materials with chains of spin-½ $Cu^{2+}$ cations.[6-10] By comparing the measured thermomagnetic properties with theoretical calculations, the role of interchain interactions $J_i$, anisotropy $\Delta$, and other sub-leading interactions can be inferred.[11-13, 10]

Spin-ladders can be thought of as an intermediate system between purely 1D spin-chains and 2D spin-sheets.[14, 15] The correlated physics is more intricate than for spin-chains without the full complexity of a 2D magnetic lattice. $Cu^{2+}$ based spin-ladders are typically formed from coplanar corner-sharing $CuO_4$ squares so that both $J_1$ and $J_2$ are AFM. The NN interactions can be broken down into those along the rungs, $J_{1,rungs}$, and those along the spines, $J_{1,spines}$. By considering the interplay between these competing interactions, a complete magnetic phase diagram can be constructed.[16-18] Other 2-rung ladder systems have been reported where the NN neighbor interactions are FM; however in these systems the AFM NNN interactions dominate.[19-21] Recently the magnetic properties of $Ba_3Cu_3In_4O_{12}$ (In-334) were reported.[22] The spin-½ $Cu^{2+}$ cations in In-334 form chains that propagate along the $c$-direction of the tetragonal unit cell and form an unusual half-twist ladder arrangement (Figure 1a and Figure 1b). In these half-twist ladders the FM $J_1$ dominate the 1D intrachain exchange, and $|J_1| \gg |J_2|$ providing access to a different part of the generalised spin-ladder phase diagram. Here we report a comparison of the magnetic properties of In-334 and its isostructural analogue, $Ba_3Cu_3Sc_4O_{12}$ (Sc-334).[23] We demonstrate the effect of altering the perovskite framework on the properties and from our thermomagnetic measurements elucidate their magnetic phase diagrams. Both materials feature three magnetic ordering regimes prior to field induced FM ordering at $\mu_oH_{SAT} \leq$ 8 T. We find that the FM $J_1$ interactions play a significant role in the magnetic ordering of these half-twist spin-ladders.

## 2. Experimental

Polycrystalline samples of In-334 and Sc-334, $Ba_3Cu_3In_4O_{12}$ and $Ba_3Cu_3Sc_4O_{12}$, were prepared using a solid-state synthesis route. Stoichiometric mixtures of pre-dried $BaCO_3$ (Alfa

Aesar, 99.95 %), CuO (Alfa Aesar, 99.7 %) and either $Sc_2O_3$ (Rare Earth Products Ltd., 99.99 %) or $In_2O_3$ (Alfa Aesar, 99.9 %) were intimately mixed. Pellets of material were then fired in air. In-334 was prepared by heating at 850 °C for 24 hours, heating 4 times at 900 °C for 24 hours and a single 24 hour heating at 950 °C while Sc-334 was made by firing at 850 °C for 24 hours and then twice at 950 °C for 24 hours. After each heating, the samples were ground into a fine powder and repressed into pellets. The progress of the reactions was monitored using a Bruker D8 Focus X-ray diffractometer operating with Cu K$\alpha$ radiation and a graphite diffracted beam monochromator. High resolution data for quantitative analysis of the crystal structure were collected over the angular range $5 \leq 2\theta \leq 90°$ with $\Delta 2\theta = 0.04°$. Rietveld analysis[24] was carried out using the *GSAS*[25] suite of programs with the *EXPGUI* interface.[26] Backgrounds were fit using a Chebyshev polynomial of the first kind and the peak shape modeled using a pseudo-Voigt function.

Thermomagnetic measurements were made using a Quantum Design Physical Property Measurements System (PPMS). Magnetic susceptibility measurements, $2 \leq T \leq 300$ K, were made after cooling in zero field (ZFC) in a range of fields up to 9 T. Isothermal magnetization measurements, $0 \leq \mu_o H \leq 9$ T, were made at selected temperatures after cooling in zero field. Specific heat measurements were made, $2 \leq T \leq 30$ K, on cooling in a range of fields; to increase the thermal conductivity, samples for these measurements were mixed with silver and pressed into dense pellets. The contribution of silver to the specific heat was then deducted.[27]

**3. Results and Discussion**
*3.2. Structure*

The structure of the In-334 and Sc-334 phases has previously been reported.[23, 28] Refinement of the current high resolution X-ray diffraction data indicated the formation of the 334 phase in our samples and show no deviations from the reported structures. In the case of In-334, a small amount of non-magnetic $In_2O_3$ (1.4(2) wt%) is observed in the refinement. This represents no significant change in the stoichiometry of In-334. The fitted X-ray diffraction data is presented in Figure 2 for In-334 and a polyhedral model of the structure is shown in the inset. The perovskite framework is highlighted in the upper right unit cell. As expected, an expansion of the unit cell occurs when the smaller $Sc^{3+}$ cations ($a$ = 11.8898(5) Å, $c$ = 8.3908(5) Å) are replaced with larger $In^{3+}$ cations ($a$ = 12.1107(3) Å, $c$ = 8.5077(2) Å).

The 334 phase can be considered as a modified perovskite phase in which ¼ of the Ba A-cation sites are removed. The resulting A-site vacancies, which have square faces, are occupied by $CuO_4$ squares on ⅔ of the faces, i.e. each Ba is replaced by three Cu atoms. Each $CuO_4$ square shares corner oxygen with adjacent $CuO_4$ squares. The A site vacancies containing these squares order crystallographically, forming chains along the *c*-axis. The orientational ordering of the $CuO_4$ squares results in a half-twist ladder arrangement of Cu ions (Figure 1a and Figure 2 inset). Due to the 1D ordering of Ba vacancies, the 334 phase has tetragonal symmetry, however, the changes in the lattice parameter relative to the parent cubic perovskite phase are very small. The expanded cell of the 334 phase is related to cubic perovskite by $a_{334} = b_{334} = 2\sqrt{2}a_c$ and $c_{334} = 2a_c$, and the resulting tetragonal distortion is quantified by $a_{334}/\sqrt{2}c_{334}$. This ratio is 1.007 and 1.002 for In-334 and Sc-334 respectively. The increase in the tetragonal distortion in In-334 represents a larger expansion in the *a* and *b* parameters compared to *c*. The compression of the *c*-axis in In-334 can be seen in an increase in the distortion of the Cu2 square-plane $\delta_{Cu2-O} = |(Cu2-O1)-(Cu2-O3)|/\langle Cu2-O\rangle$ in In-334 ($\delta_{Cu2-O}$ = 0.03 and 0.01 for In-334 and Sc-334 respectively). Unexpectedly given the large change in the lattice parameters between In-334 and Sc-334, no significant changes to the Cu-O bond lengths is observed. This implies that the structural changes are due to the change in ionic radius of the B-cations and primarily affect the perovskite part of the structure. The only significant change in the Cu environment is the Cu1-O1-Cu2 bond angle, which is reduced in Sc-334 (87.8°) and corresponds to a compression of the half-twist ladder as a result of chemical pressure from the Ba-Sc-O framework. For In-334 the half-twist ladders are fully extended (Cu1-O1-Cu2 = 90.7°) to fill the increased volume of the vacant Ba sites.

### *3.2. Magnetic Characterization*

The ZFC magnetic susceptibility, $\chi = M/H$, of In-334 in selected fields is shown in Figure 3a. In low fields, $0.1 \leq \mu_oH \leq 4$ T, a sharp cusp characteristic of 3D AFM order is observed at a temperature that we now denote as $T_N$. $T_N$ decreases from 12.3(2) K to 6.0(5) K upon increasing the field from 0.1 to 4 T (data not shown). Above 4 T the AFM ordering transition is absent and $\chi$ continues to increase on cooling. At high temperatures $\chi$ is field independent, and fits to the Curie-Weiss law, $(\chi-\chi_0) = C/(T-\theta)$, were carried out using the data collected at 1 T for $T > 150$ K, where M vs. H is linear. The effective magnetic moment inferred from the fit is consistent with spin-½ $Cu^{2+}$, $\mu_{eff}$ = 2.0 $\mu_B$ per Cu, whilst the Weiss temperature, $\theta$ = 56 K, indicates a preponderance of FM interactions.

Isothermal magnetization measurements of In-334 at selected temperatures are shown in Figure 4a. Below $T_N$ the isothermal magnetization has an 'S' shaped curvature and saturates at relatively low fields, $\mu_o H_{SAT} = 5.2$ T at 2 K. The saturation magnetization, $M_{SAT} = 1.05$ $\mu_B$ per Cu, is consistent with a field induced FM state with spin-½ $Cu^{2+}$ cations with $g_{av} = 2.1$ ($M_{SAT} = gS$). Differentiation of the isothermal magnetization curve (Figure 4b) indicates two spin re-orientation transitions below the saturation field for $T < T_N$. The lower field transition is sharp, indicating an actual phase transition. On the other hand the higher field anomaly is more diffuse and could be either an anisotropy broadened phase transition or simply a cross-over between regimes with different correlations. Both anomalies shift to lower fields with increasing temperature and are only observed in $dM/d\mu_o H$ for $T > T_N$.

The specific heat, $C_p$, of In-334 at selected fields is shown in Figure 5. At $T_N$ a sharp peak in both $C_p$ and $C_p/T$ is observed. This peak corresponds to the 3D AFM ordering transition observed in the magnetic susceptibility and decreases in temperature on application of a field. No evidence for the spin re-orientations indicated by the isothermal magnetization data is observed in $C_p$ at high fields, from which we infer that they do not involve a substantial change in entropy. At lower temperatures, $T < 8$ K, a broad Schottky anomaly is observed which, within the resolution of our measurements, is field independent. As a field is applied there is an overall reduction in $C_p/T$ and hence a reduction of the magnetic entropy observed in our measurements. This most likely arises from magnetic entropy being pushed to low temperature features in the applied field, $T < 2$ K, but could be due to enhanced high $T$ magnetic fluctuations $\mu_o H > 0$.

The results from our thermomagnetic measurements on In-334, reported above, are in agreement with those reported elsewhere.[22] In the following discussion we will compare them with isostructural Sc-334 and present an analysis of the properties for $T > 2T_N$, based on a modified 1D FM-AFM spin-chain model.

The magnetic susceptibility of Sc-334 shown in Figure 3b is qualitatively similar to that of In-334. However, the 3D AFM ordering transition occurs at a higher temperature, $T_N = 15.2(1)$ K at $\mu_o H = 0.1$ T, and persists to higher fields, $\mu_o H \leq 7$ T. At high temperatures, fits to the Curie-Weiss law are consistent with spin-½ $Cu^{2+}$, $\mu_{eff} = 2.1$ $\mu_B$ per Cu and dominant FM interactions, $\theta = 52$ K.

Isothermal magnetization measurements for Sc-334 are shown in Figure 4c. Below $T_N$ Sc-334 shows the same 'S' shaped curvature observed in In-334 but the higher saturation field 'stretches' the curve. At 2 K, Sc-334 saturates to $M_{SAT} = 1.0$ $\mu_B$ per Cu ($g_{av} = 2.0$) at $\mu_o H_{SAT} = 8$

T. The derivative of the isothermal magnetisation measurements, Figure 4d, indicates two distinct features for $H < H_{SAT}$. The lower field peak is sharp and shifts from 2.4 T at 2 K to a weak feature at 2 T at 15 K. The higher field anomaly is significantly broader and decreases more dramatically in field as the temperature is increased, so that at 12 K only a weak feature at 4.2 T remains.

A decrease in $T_N$ with increasing applied field is consistent with specific heat for Sc-334 (Figure 5). Again, no evidence for field driven transitions is obtained from the specific heat. At lower temperatures, $T < 8$ K, a broad Schottky anomaly is observed which, within the resolution of our measurements, is field independent. The magnitude of the Schottky anomaly in Sc-334 appears to be smaller than for In-334. As for In-334, there is an overall reduction in $C_p/T$ with increasing applied field, indicating a reduction in the change in the magnetic entropy, $\mu_o H < 0$, within the temperature range probed.

The low temperature magnetic behaviors of In-334 and Sc-334 are similar; both appear to develop 3D AFM for $T < 20$ K, and have two field-driven anomalies prior to forming a saturated field-induced FM state. We have summarized this in the tentative magnetic phase diagrams shown in Figure 6. The two magnetic phase diagrams are qualitatively similar. In addition to the reduction of $T_N$ and $H_{SAT}$ in In-334, the transitions to the intermediate magnetic phases are also shifted to lower $T$ and $H$ in In-334. To explore the nature of the magnetic phases, neutron diffraction data or direction-dependent single-crystal measurements of the thermomagnetic properties are required. The zero-field and field induced FM ordering have already been investigated in Sc-334 where a change in the ordering regime has been observed.[29, 30] To date, no definitive refinement of the magnetic structure has been obtained for either the zero field or $H > H_{SAT}$ regime, although a zero field ground state has been proposed for In-334.[22]

*3.3 Magnetic Modeling*

The nature of the magnetic interactions in 334 is governed by the half-twist spin-½ ladders that propagate along the *c*-axis of the unit cell. In this novel arrangement there are two distinct Cu polyhedral sites, Cu1 and Cu2, positioned parallel and perpendicular to the *xy*-plane respectively (Figure 1a). This results in 3 sets of mutually orthogonal easy axes for the $CuO_4$ squares. The $CuO_4$ squares share corners and the positions of the Cu ions within the vacant A sites results in <Cu-O-Cu angles close to 90° and FM $J_1$ for all of the NN interactions. When considering the longer range intrachain interactions we assume their strength is inversely related to the number of mediating oxygen atoms; this results in a slightly different picture of the magnetic exchange than previously proposed.[22] While a single AFM $J_2$ (Cu2-O-O-Cu2) is mediated by two oxygen atoms (Figure 1b), two interactions involve 3 oxygen atoms, the AFM Cu1-Cu1 exchange, $J_3$, and a second Cu2-Cu2 exchange, $J_3$'. These interactions involve the same number of oxygen atoms as the AFM interchain coupling, $J_i$ (Cu2-Cu2), while a second interchain coupling, $J_i$' (Cu1-Cu1), involves 4 mediating oxygen ions. Determining which of these interactions is significant in the magnetic ground state is non-trivial and would involve a complex theoretical analysis.

By neglecting all but the NN and NNN interactions it is possible to makes inferences about the $J_1$ and $J_2$ interactions by using a modified analysis derived from 1D spin-½ chains. Extensive theoretical work has been carried out on the spin-½ FM-AFM frustrated chains and a complete magnetic ground state derived.[3, 13, 12, 4, 11] In these frustrated chains, as in the 334 phase, the $J_1$ interactions are FM and the $J_2$ are AFM. The behavior of the FM-AFM system has been determined as a function of $\alpha = J_2/J_1$ and a transition from a short-range ordered (SRO) regime ($\alpha < \alpha_c$) to FM ($\alpha > \alpha_c$) is predicted at a quantum critical point (QCP) when $\alpha_c$ = -0.25.[31, 32] In materials close to the QCP, such as LiCuSbO$_4$[10] and Rb$_2$Cu$_2$Mo$_3$O$_{12}$,[7] the thermomagnetic properties observed are similar to those of In-334 and Sc-334; specifically, the shape of the '*S*' curvature in the isothermal magnetization and the decrease in the ordering temperature with increasing field.

Given the similarity between the thermomagnetic properties of the spin-½ spin-chains and the half-twist ladders in 334, we consider in our model the interplay of only the $J_1$ and $J_2$ intrachain interactions described above and shown in Figure 1b. Although quantitative modeling requires considering interchain exchange interactions, a purely 1D analysis may be appropriate for $T > 2T_N$ where the contributions from interchain interactions are less important. We obtained the magnetic properties by exact diagonalization (ED) for finite chain with N = 3n spins (n ≤ 6), isotropic $J_1 < 0$ and $J_2 = \alpha J_1 > 0$, and periodic boundary conditions:

$$H(\alpha)/|J_1| = -\sum_{\langle n,n'\rangle} \vec{s}_n \cdot \vec{s}_{n'} - \alpha \sum_{\langle n,n''\rangle} \vec{s}_n \cdot \vec{s}_{n''} - h\sum_n s_n^z \qquad \text{Eq. (1)}$$

Here h = g$\mu_B$B/|J$_1$| is the Zeeman interaction with a magnetic field B and g is the average isotropic g-tensor, $g_{av}$ = 2.1. The exchange interactions considered are shown in Figure 1b, <n, n'> spin pairs correspond to the $J_1$ exchange between Cu1 and its four NN Cu2, whereas <n, n"> spin pairs correspond to $J_2$ exchange between Cu2 and its four NNN Cu2. Neither the anisotropy of the exchange interactions nor the g-tensor anisotropy is included. The calculated magnetic susceptibility and isothermal magnetization for N = 18 are compared with experimental data for In-334 in Figure 7a and Figure 7b. Essentially identical fits are found for parameters ranging

from $J_1$ = -125 K, $\alpha$ = 0 to $J_1$ = –155 K, $\alpha$ = –0.06. The corresponding Sc-334 fits obtained with parameters ranging from $J_1$ = -105 K, $\alpha$ = 0 to $J_1$ = –140 K, $\alpha$ = –0.08 are shown in Figure 7c and Figure 7d. Consistently, all of the fits have $|J_1| \gg |J_2|$ and we find that the ground state is always FM. In all models the $J_1$ interactions are more FM in In-334, this is consistent with the <Cu-O-Cu bond angle in In-334 being closer to 90º and In-334 having a higher Weiss temperature. The fit to $\chi$ is reasonable down to temperatures of ~ $2T_N$; at T < $2T_N$ the increasing contributions of the AFM interchain exchange interactions suppresses the measured $\chi$. Similarly, the model for the isothermal magnetization in In-334 at 20 K ($2T_N$ = 25 K) deviates from the observations at low field where AFM $J_i$ and $J_i'$ interactions become important. The inset of Figure 7a and Figure 7c show finite-size effects at $\alpha$ = 0. The N = 15 and 18 curves are nearly identical, whilst the N = 9 and 12 curves suggest rapid convergence towards the infinite chain limit.

The $J_2$ interactions in In-334 and Sc-344 are apparently sufficiently weak to allow the spin-chains to be modeled using a single FM $J_1$ NN interaction; this is unique within oxide spin-ladder compounds which are all found to be AFM. In other 2-leg spin-ladder systems, such as $SrCu_2O_3$ (Figure 1c), the corner sharing networks of $CuO_4$ squares have <O-Cu-O ~ 180º and both the $J_1$ and $J_2$ interactions are AFM (Figure 1d). Whilst this can also lead to magnetic frustration and competing ground states, FM order is not favored. Consideration of the magnetic interactions in the 334 phase shows that whilst the FM $J_1$ connects adjacent Cu1 and Cu2 atoms within the chains, the AFM $J_2$ coupling is only between the Cu2 atoms along the chain. The AFM coupling between the Cu1 atoms and the Cu2 atoms (across the chain) are described by extended, weaker $J_3$ and $J_3'$ interactions. Therefore, unlike in other 2-leg ladder systems, the AFM extended intrachain interactions are suppressed by the connectivity of the $CuO_4$ squares and the FM $J_1$ intrachain interactions is for once dominant.

The model described above, which includes only intrachain interactions, cannot account for the sharp transition observed at $T_N$ in both the magnetic susceptibility and specific heat. These features are clearly indicative of 3D ordering and imply that the $J_i$ interactions are significant in the formation of a 3D ordered state. The difference in $T_N$ and $H_{SAT}$ for the two 334 phases can be qualitatively explained by considering the increase in the interchain separation in In-334. Thus, in In-334 the $J_i$ interactions are weaker and only become significant at lower $T$ and $H$. The complexity of the magnetic ordering, with a number of distinct magnetic transitions, suggests additional interactions may also come into play. These could be the AFM intrachain interactions, $J_2$, $J_3$ and $J_3'$, or may be due to anisotropic terms within the Cu-ladders. To explore the behavior, $T < 2T_N$, the magnetic exchange interactions must be determined either by theoretical calculations and subsequent fitting to the thermomagnetic data and/or single crystal experiments to measure the ferromagnon dispersion relation in the fully magnetized state.[33]

## 4. Conclusion

The half-twist ladder arrangement of the In-334 and Sc-334 phases is unique within the family of 2-leg ladders in that it has dominant FM $J_1$ intrachain interactions. At low fields the extended AFM interactions dominate, resulting in 3D AFM ordering. As the field increases, two further magnetic regimes are observed prior to field driven FM ordering at $H_{SAT}$. $H_{SAT}$ for both In-334 and Sc-334 is readily accessible using conventional superconducting magnets allowing experimental access to all of the magnetic regimes to be readily investigated. We speculate that, in addition to manipulation of the properties by chemical substitution, it may be possible to tune the onset of 3D AFM ordering by the application of pressure.

## 5. Acknowledgments

The authors acknowledge helpful discussions with Shuang Jia. This research was supported by the U.S. Department of Energy, Office of Basic Energy Sciences, Division of Materials Sciences and Engineering under Award DE-FG02-08ER46544.

Figure 1: Polyhedral models of the Cu ladders in (a) the half-twist ladder 334 phases and (c) the 2-leg ladder material $SrCu_2O_3$. The $J_1$ and $J_2$ interactions are shown in (b) and (d) for the 334 phases and $SrCu_2O_3$ respectively.

Figure 2 (color online): Observed (o) and calculated (-) X-ray diffraction data for In-334 collected at 298 K. The difference curve is also shown; reflection positions are indicated by the vertical lines for the main phase (upper), and the $In_2O_3$ impurity phases (lower). A polyhedral model looking down the c-axis chains is shown in the inset. The unit cell on the upper left includes the $InO_6$ polyhedra to highlight the perovskite-based framework in which the $Cu^{2+}$ half-twist ladders are contained.

Figure 3: ZFC Magnetic susceptibility, $\chi = M/H$, in selected fields for (a) In-334 and (b) Sc-334. The inverse susceptibility, $(\chi-\chi_o)^{-1}$, in a 9 T field is shown in the inset, the Curie-Weiss fit, T > 150 K, is also shown.

Figure 4: Isothermal magnetisation of (a) In-334 and (c) Sc-334 at selected temperatures. The derivative, $d\mu_oH/dT$, is shown in the (b) and (d) for In-334 and Sc-334 respectively.

Figure 5: $C_p/T$ as a function of temperature for (a) In-334 and (b) Sc-334. $C_p$ as a function of $T$ is shown in the inset.

Figure 6 (color online): Proposed magnetic phase diagrams for (a) In-334 and (b) Sc-334 as a function of field, $\mu_0H$, and temperature, $T$, derived from the thermomagnetic measurments. $T_N$ obtained from the magnetic susceptibility and specific heat is shown as blue squares and green triangles respectively. $H_{SAT}$ and the position of the two features in $d\mu_oH/dT$ observed are shown as black squares, red circles and light blue stars respectively.

Figure 7 (color online): Fits using a modified FM-AFM spin Hamiltonian (Eq. (1)) to the magnetic susceptibility, $\chi = M/H$, and isothermal magnetization of (a-b) In-334 and (c-d) Sc-334. Fits from 18 spins and $\alpha = 0$, $J_1 = -125$ K are shown in green and $\alpha = -0.06$, $J_1 = -155$ K in blue for In-334. For Sc-334, fits from 18 spins and $\alpha = 0$, $J_1 = -105$ K are shown in green and $\alpha = -$ 0.08, $J_1$ = -150 K in blue. The effect of the size of the spin-chains on the modelled magnetic susceptibility is inset in (a + c) for $\alpha = 0$ and the appropriate $J_1$.

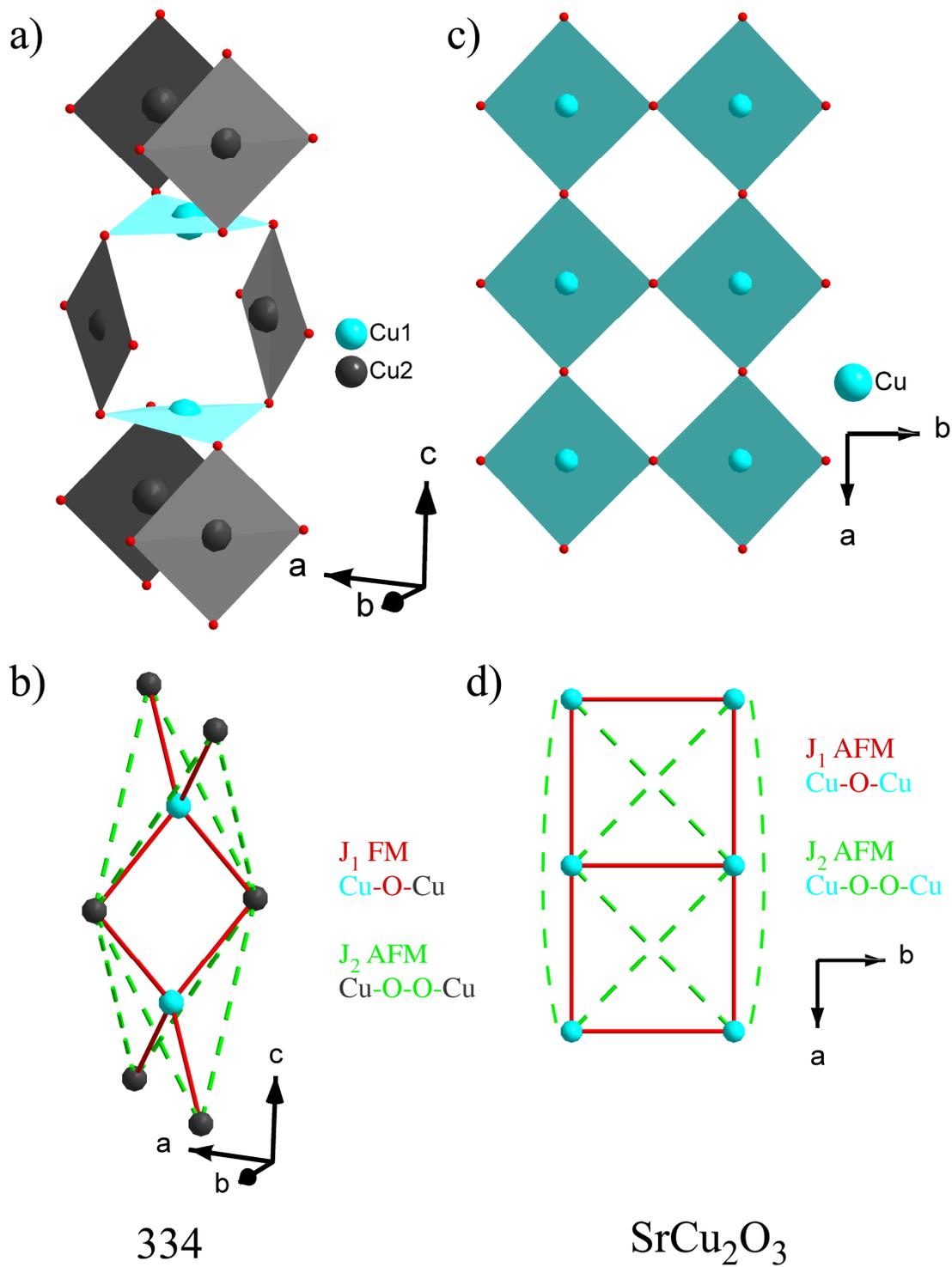

Figure 1

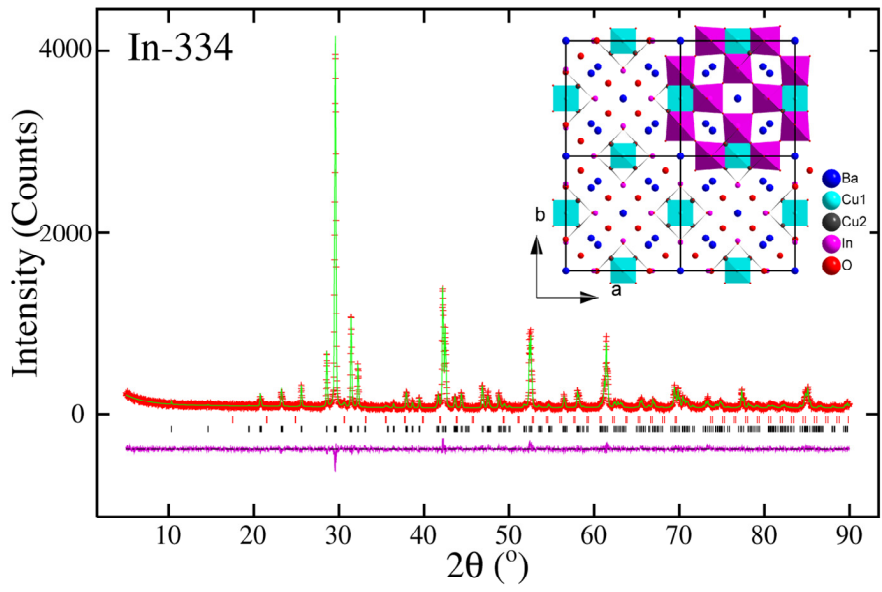

Figure 2

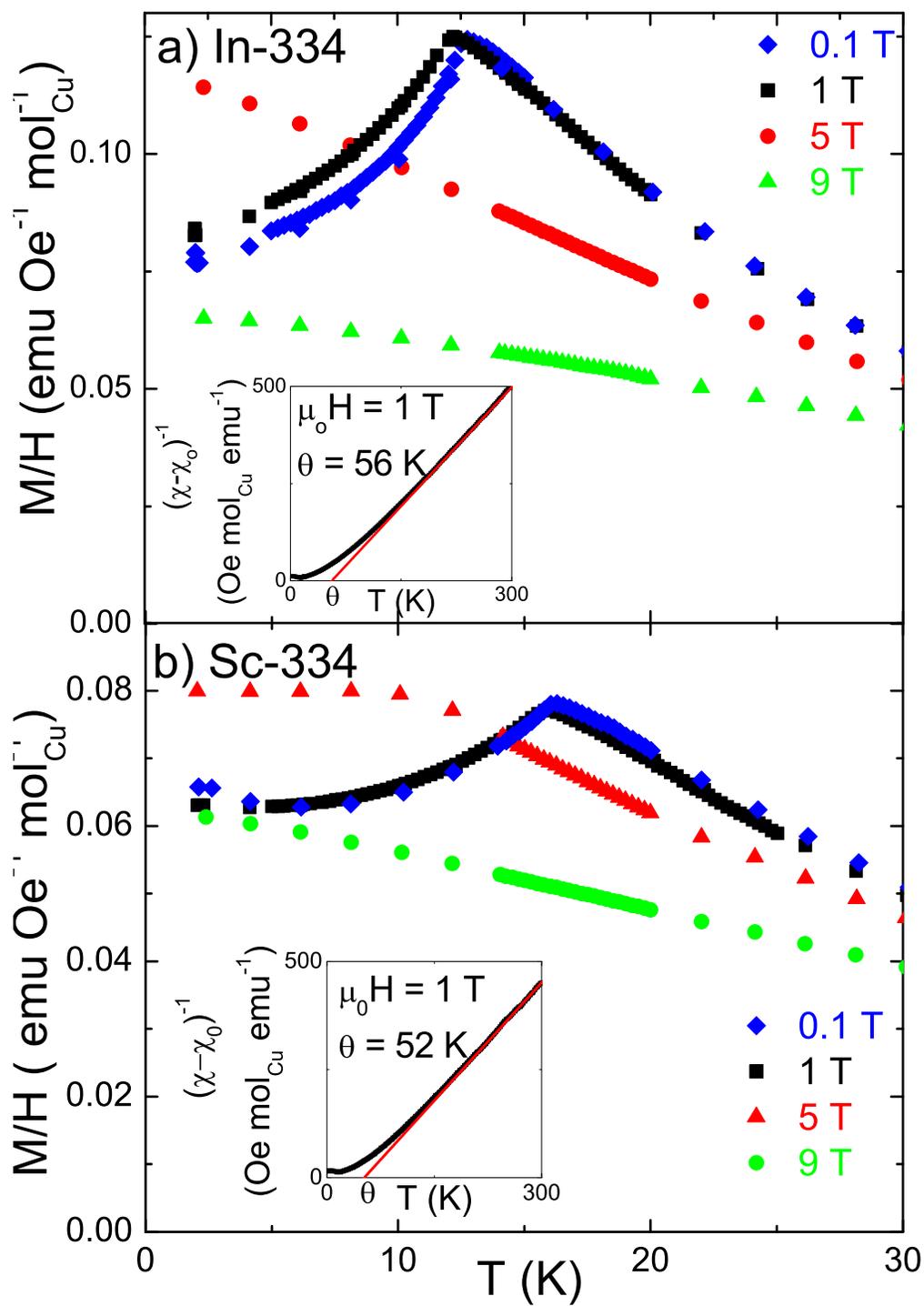

Figure 3

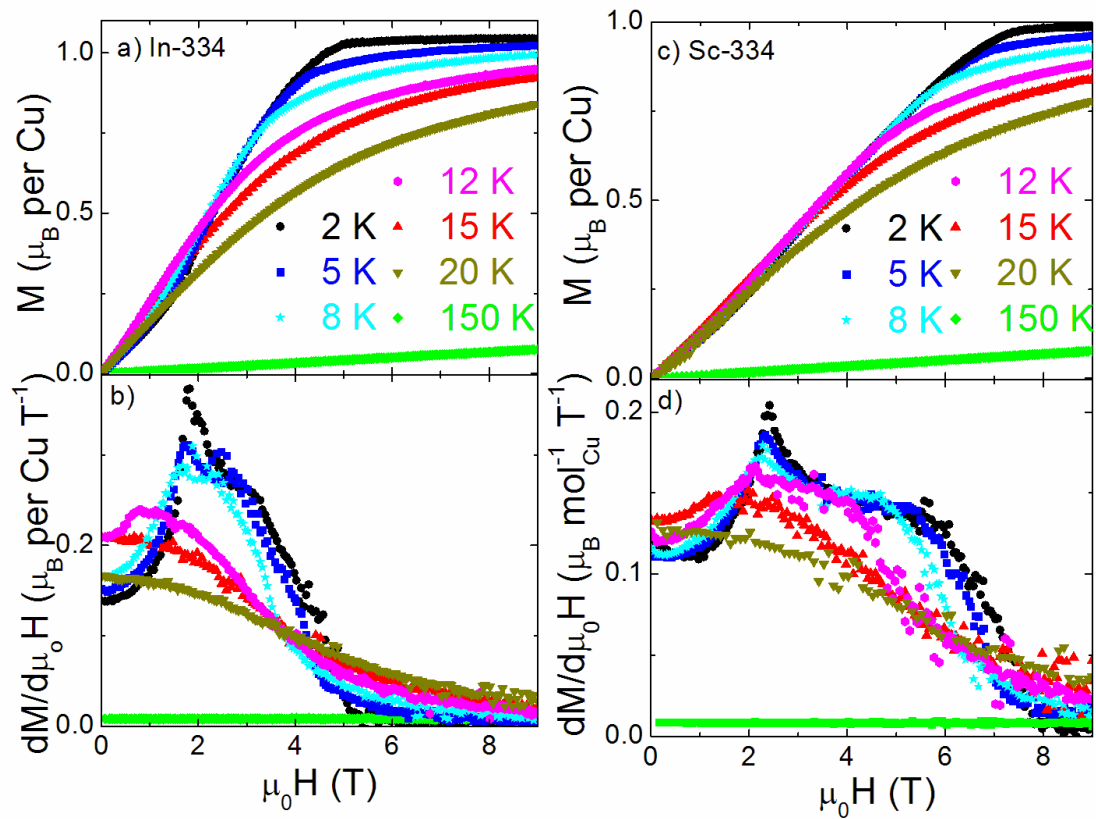

Figure 4

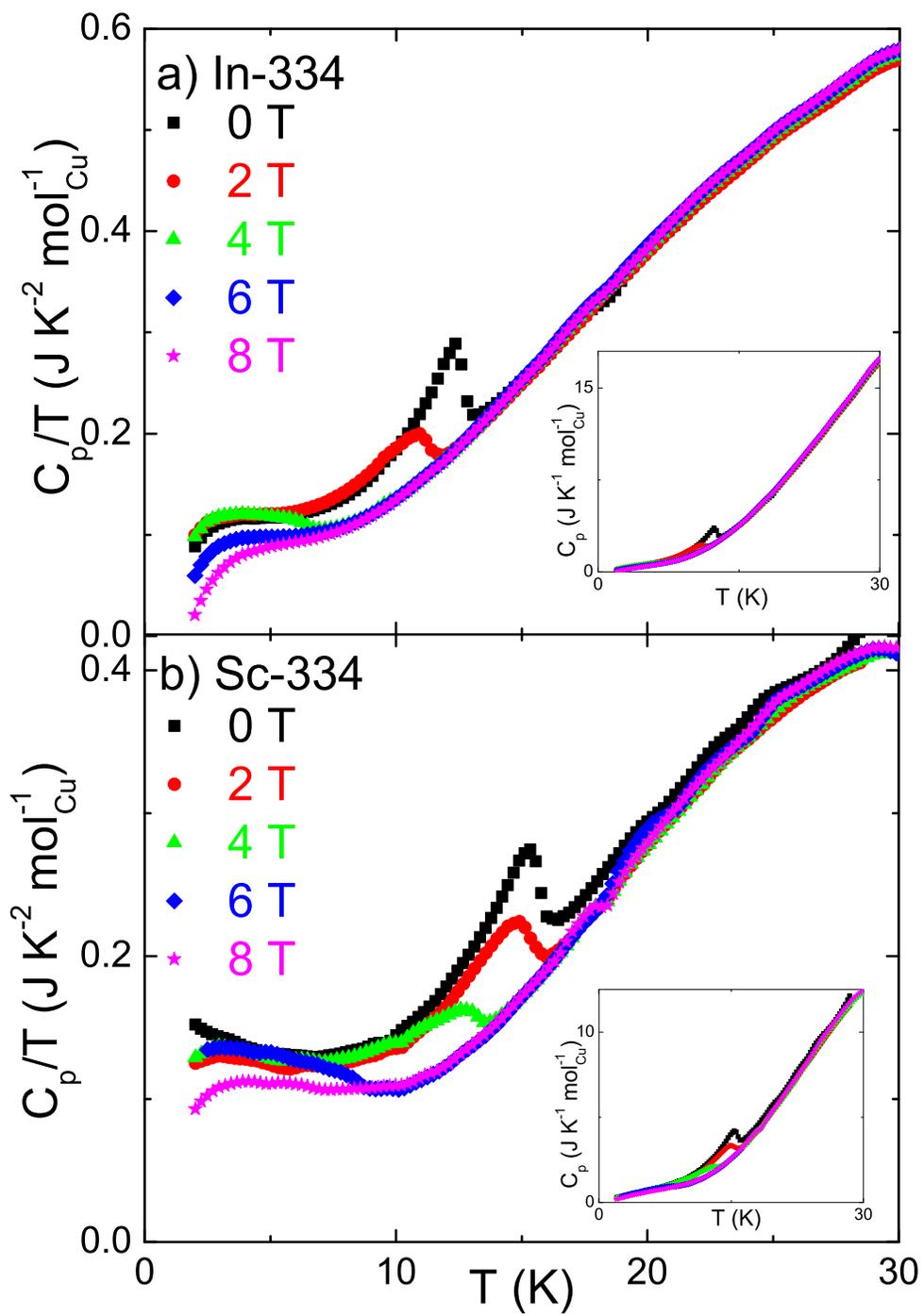

Figure 5

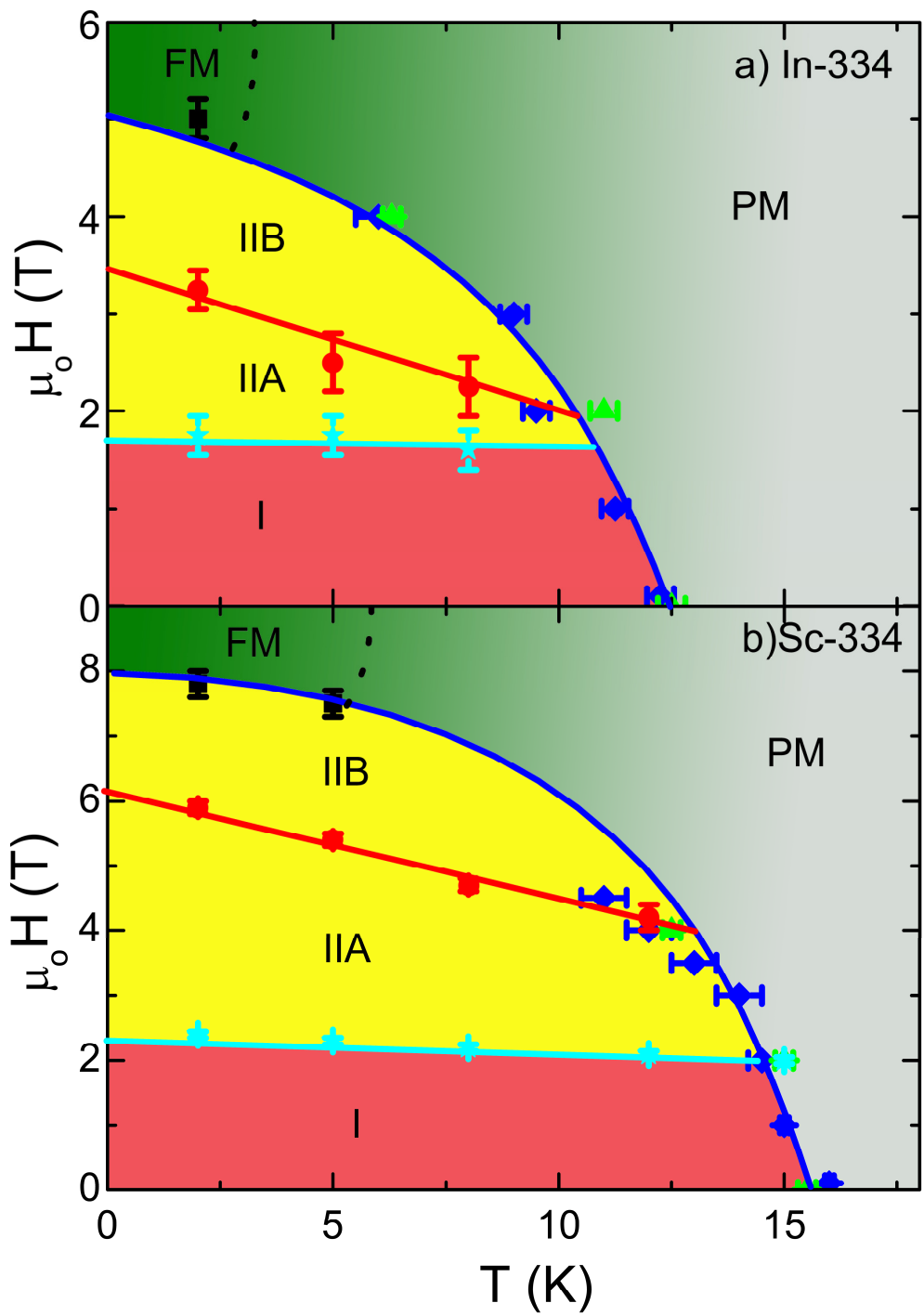

Figure 6

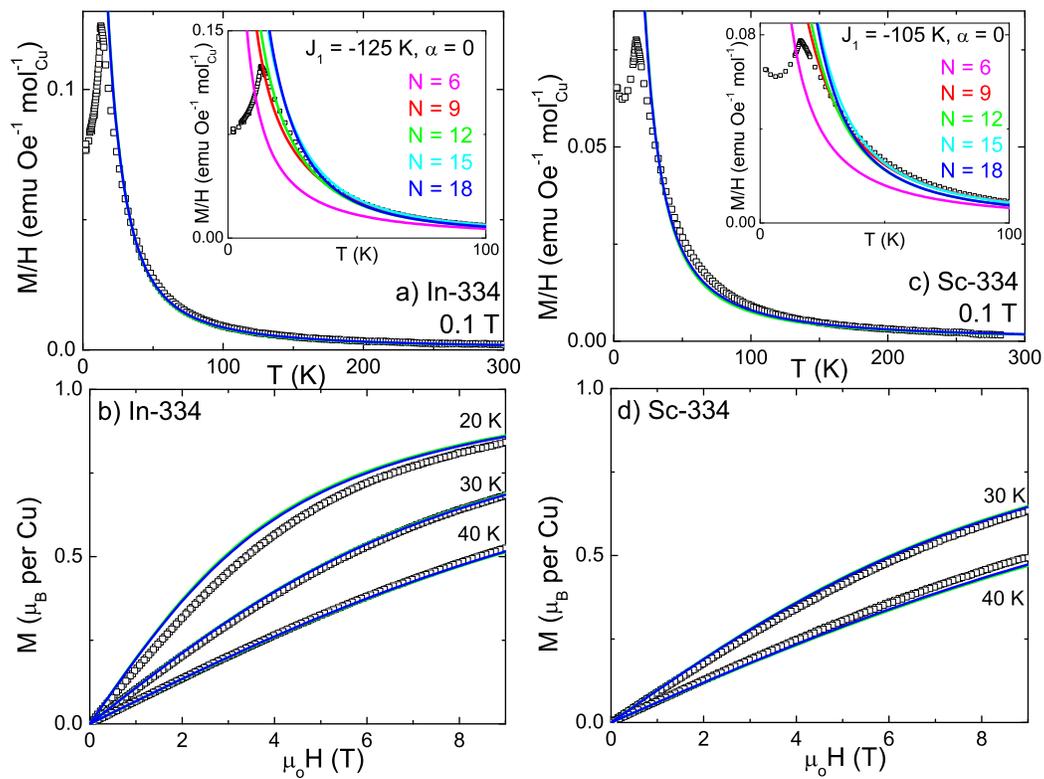

Figure 7